\documentclass[aps,a4paper,superscriptaddress,twocolumn]{revtex4-2}
\usepackage{tensor}
\usepackage{graphicx}
\usepackage{amsmath}
\usepackage{amssymb}
\usepackage{enumerate}
\usepackage{subfigure}
\usepackage{tabularx}
\usepackage[colorlinks=true, pdfstartview=FitV, linkcolor=blue, citecolor=red, urlcolor=black, breaklinks=true]{hyperref}
\newcommand{\be}{\begin{equation}}
\newcommand{\ee}{\end{equation}}
\newcommand{\ben}{\begin{eqnarray}}
\newcommand{\een}{\end{eqnarray}}
\newcommand{\bes}{\begin{subequations}}
\newcommand{\ees}{\end{subequations}}
\def\bal#1\eal{\begin{align}#1\end{align}}

\newcommand{\sech}{{\rm sech}}

\begin{document}
\title{Manipulating the internal structure of Bloch walls}
\author{D. Bazeia}%\email{dbazeia@fisica.ufpb.br}
\affiliation{Departamento de F\'\i sica, Universidade Federal da Para\'\i ba, 58051-970 Jo\~ao Pessoa, PB, Brazil}
\author{M.A. Marques}%\email{marques@cbiotec.ufpb.br}
\affiliation{Departamento de Biotecnologia, Universidade Federal da Para\'\i ba, 58051-900 Jo\~ao Pessoa, PB, Brazil}\affiliation{Departamento de F\'\i sica, Universidade Federal da Para\'\i ba, 58051-970 Jo\~ao Pessoa, PB, Brazil}
\author{M. Paganelly}%\email{matheuspaganelly@gmail.com}
\affiliation{Departamento de F\'\i sica, Universidade Federal da Para\'\i ba, 58051-970 Jo\~ao Pessoa, PB, Brazil}
\begin{abstract}
In this work, we describe a procedure to manipulate the internal structure of localized configurations of the Bloch wall type. We consider a three-field model and develop a first order formalism based on the minimization of the energy of the static fields. The results show that the third field may be decoupled and used to change the geometric arrangement of the Bloch wall, giving rise to a diversity of modifications of its internal structure. 
The procedure captures effects that goes beyond the standard situation and can be used in several applications of practical interest, in particular, for the study of the magnetization of magnetic materials at the nanometric scale.
\end{abstract}
%\date{\today}
\maketitle 

\section{Introduction}

In high energy physics, kinks are localized finite energy topological structures that appear in models described by a single real scalar field in two-dimensional spacetime. When immersed in higher dimensions, kinks become domain walls, and both have been studied for more than fifty years and have many interesting applications in several areas of nonlinear science, for instance, in biological molecules, in Bose-Einstein condensates, in conducting polymers, in ferroelectric elements, in magnetic materials, in optical fibers and in other systems; see, e.g., Refs. \cite{A,A1,B,C,D,OS,book,E} and references therein. 

The existence of localized structures may be related to the presence of phase transitions in the system under investigation. This is of particular interest during the cosmic evolution of the Universe, and here we recall the recent work aimed to review the basics to understand phase transitions in the early universe, their significance and experimental signatures \cite{REV}. In condensed matter, localized structures find applications of practical interest in several distinct contexts, among the many possibilities we may suggest the recent review on localized structures in fiber lasers such as bright and dark solitons and their compositions as bright-bright, bright-dark and dark-dark configurations \cite{FL}, the studies of the double-well energy landscape in ferroelectric materials \cite{N,JAP} and of domain wall skyrmions in magnetic systems \cite{PhysRevB,NC}, and references therein. Another possibility of current interest concerns the study of multiferroic materials \cite{N1,N2}, which support composite domain wall configurations where ferroelectric polarization and magnetization are successfully controlled by magnetic and electric fields. 

One of the simplest kinks or domain walls that appears in high energy physics is controlled by a single real scalar field. Usually, kinks have the profile of the hyperbolic tangent, but we can also have models describing lumps, which are bell-shaped field configurations with the profile of the hyperbolic secant. In general, kinks are stable field configurations that attain topological information, but lumps are unstable and so of minor interest in high energy physics. In optical fibers, however, the electric field of the laser beam can be related to the square of the field configuration and so, kinks and lumps can be used to represent dark and bright solitons, respectively. In magnetic materials, kinks may describe the magnetization and in the simplest case of a single field, it is known as N\'eel wall. However, if one adds another real scalar field, it is possible to construct kinks with internal structure. In this case, the domain wall is usually known as Bloch wall. The motivation follows as in Ref. \cite{PRE0}, but here we concentrate on the Bloch wall, in which the second field may describe another degree of freedom and find other applications of current interest. For instance, the second field may control the thickness of the Bloch wall, which, in the case of magnetic systems, may be related to the anisotropy, exchange and magnetic energy parameters that control the Bloch wall profile. This is an issue that motivates the present investigation and we revisit the subject in Sec. \ref{sec2}, where we briefly review the two possibilities to construct analytical solutions for both the N\'eel and Bloch walls.

The present investigation considers relativistic scalar fields and we focus mainly on the presence of static configurations, so the dynamical evolution of the fields is not visible and the study is similar to the case of nonrelativistic fields. Other approaches are also possible, aimed to discuss distinct issues. One can, for instance, consider relativistic scalar field models to investigate the behavior of fermions in the background of a kink \cite{Jackiw1}, its relation with localized structures in polyacetylene \cite{Jackiw2}, the presence of topological twistons in crystalline polyethylene \cite{Ventura} and also, to describe excitations of a long graphene nanoribbon buckled in the transverse direction in the presence of kink and kink-antikink pair \cite{PRBkink}. One can also consider the nonlinear Schr\"odinger, Gross-Pitaevskii, Ginzburg-Landau, Landau-Lifshitz, Landau-Lifshitz-Gilbert or even other equations; in particular, in Ref. \cite{Arge} the authors dealt with magnetic domain wall motion considering an nonrelativistic scalar field model to describe the effective motion of the magnetization. We notice, in particular, that the scalar field self-interaction considered in \cite{Arge} is also of the fourth-order power, well-known in high energy physics and in other areas of nonlinear science and similar to the one to be considered in the present work.

Another motivation for developing the present study comes from the recent investigation \cite{multikink}, in which one described a novel effect which is caused by the presence of another scalar field, that serves as a kind of geometrical deformation of the medium where the other field evolves, because the extra field can be added to act to constrain such medium. This idea was further considered in Ref. \cite{BMM} in the presence of fermions, to describe how the fermionic degrees of freedom are affected by the presence of the geometrically constrained kink. As it was shown, the geometrical constraints induce interesting modifications in the fermionic field, which can be used to control the domain wall polarity, for instance \cite{BMM,R,L}. Related to this, other effects of manipulation of magnetic order, domain walls, and skyrmions were recently reviewed in \cite{RMP}. Some of the effects of magnetic order explored in \cite{RMP} are of direct interest to the present work, in particular, the domain wall response in the presence of current, spin transfer and spin-orbit torque.

In the present work, we focus on the possibility to manipulate the internal structure of Bloch walls. We start the investigation in Sec. \ref{sec2}, where we review the basic properties of kinks or domain walls in models described by one and two real scalar fields and chiral Bloch walls. In Sec. \ref{CBW} we describe a novel model, composed by three real scalar fields. We study the presence of localized configurations that describe Bloch walls with interesting internal structures. The procedure is based on the search for first order differential equations that solve the equations of motion and give rise to stable structures. We end the work in Sec. \ref{conclu}, where we add comments and other possible lines of investigations. 

\section{N\'eel and Bloch walls}\label{sec2}

Let us start reviewing some basic properties of the N\'eel and Bloch walls. We consider the well-known model, described by the real scalar field $\phi$ with Lagrange density
\be
{\cal L}=\frac12\partial_\mu\phi\partial^\mu\phi-V(\phi).
\ee
Here we are working in $(1,1)$ spacetime dimensions, with $x^\mu=(x^0=t,x^1=x)$, $\eta_{\mu\nu}=\textrm{diag}(1,-1)$, and $V(\phi)$ is the potential. Invariance under spacetime translation gives rise to the energy momentum tensor
\begin{equation}\label{tmunu}
    T_{\mu\nu}=\partial_{\mu}\phi\partial_{\nu}\phi-\eta_{\mu\nu}\mathcal{L}.
 \end{equation}

The model with the potential taken in the form
\be
V(\phi)=\frac12 (1-\phi^2)^2,
\ee
is known as the $\phi^4$ model. This potential is the prototype of the Higgs potential, engendering the
fourth-order power in the field and spontaneous symmetry breaking. Here we follow \cite{Bazeia} and our notation includes $\hbar=c=1$ and the time and space coordinates $t$ and $x$, and the field $\phi$ are all dimensionless.

The potential can be written in terms of the auxiliary function $W(\phi)$ in the form
\be
V=\frac12\; W^2_\phi,
\ee
where $W_\phi=dW/d\phi= (1-\phi^2)$, such that
\be\label{w}
W(\phi)=\phi-\frac13\, \phi^3.
\ee 
The equation of motion for static solutions is
\be  
\phi{''}=W_\phi W_{\phi\phi}=-2\,\phi\, (1-\phi^2),
\ee
where $\phi{''}=d^2\phi/dx^2$. In this work, we will use the prime to represent derivative with respect to the coordinate $x$. As it is well-known, the solutions \cite{bogo} also obey the first-order equations
\be
\phi'=\pm W_\phi=\pm (1-\phi^2).
\ee
The two equations are related by the change $x\to -x$, so we will deal only with the positive sign. The solution is given by $\phi(x)=\tanh(x)$, and it  represents the order parameter of the N\'eel wall, with conventional unitary amplitude and thickness, for simplicity. In this case, the energy density $\rho=T^{00}$ and stress $p=T^{11}$ are
\begin{subequations}
	\begin{align}\label{ened}
\rho &=\frac{1}{2}{\phi'}^{2}+V(\phi),\\
p &= \frac{1}{2}{\phi'}^{2}-V(\phi).
\end{align}
\end{subequations}
For the above solution we have $p=0$ and $\rho=\sech^4 (x)$, giving the total energy $E=4/3$.
As we can see from Eq. \eqref{w}, we could add two distinct parameters to the model, to control the amplitude, width and the energy of the localized solution. However, we have instead made them unity, to describe the standard configuration.

The above model can be enlarged to accommodate another scalar field, $\chi$, as considered in \cite{b1} and in other investigations, in particular in Refs. \cite{Shifman,B2,Juan,Avelino}. This is achieved with the model
\be  
{\cal L}=\frac12\partial_\mu\phi\partial^\mu\phi+\frac12 \partial_\mu\chi\partial^\mu\chi-V(\phi,\chi),
\ee
where the potential now depends on the two fields. The interesting model is described by 
\be
V=\frac12W_\phi^2+\frac12 W_\chi^2,
\ee 
where $W=W(\phi,\chi)$ has the form
\be
W=\phi-\frac13 \phi^3- r \phi\chi^2,
\ee
with $r>0$ being a real parameter. The two fields $\phi$ and $\chi$ are now used to represent a new order parameter, a vector in the $(\phi,\chi)$ plane, appropriate to describe a Bloch wall configuration with two degrees of freedom. The equations of motion for static solutions can be solved by solutions of the first-order equations
\bes
\bal
&\phi'=W_{\phi}=(1-\phi^2)- r \chi^2,\\
&\chi'=W_{\chi}=-2  r \phi\chi.
\eal
\ees

%%%%%%%%%%
	\begin{figure}[t!]
		\centering
		\includegraphics[width=8cm,trim={0cm 0cm 0 0},clip]{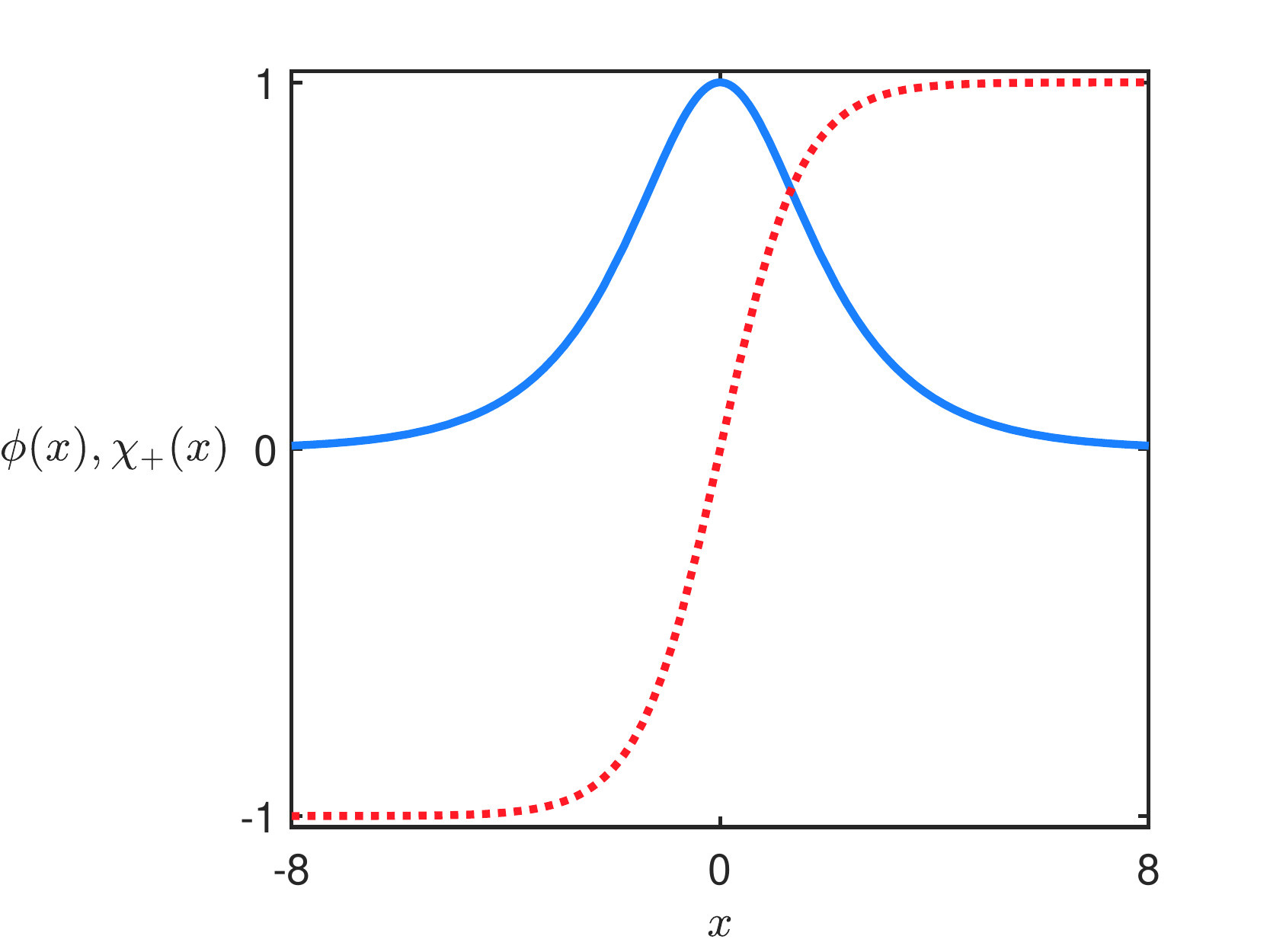}
		\caption{The solutions $\phi(x)$ (red, dotted line) and $\chi_+(x)$ (blue, solid line) in Eqs.~\eqref{111}, depicted for $r=1/3$.}
		\label{fig1}
	\end{figure}
%%%%%%%%%%

This model has interesting solutions, in particular, for $r\in (0,1/2)$ we get
\bes\label{111}
\bal
&\phi=\tanh(2rx),\label{111a}\\
&\chi_\pm=\pm\sqrt{\frac1{r}-2}\; \sech(2rx)\label{111b}.
\eal
\ees
These solutions can be found with the help of the trial orbit method \cite{Raja,bflrorbit}, which we further explain in Sec. \ref{CBW}, where we study the more complicated case involving the presence of three real scalar fields. The solutions \eqref{111} are $(\phi(x),\chi_+(x))$ and $(\phi(x),\chi_-(x))$. We depict $(\phi(x),\chi_+(x))$ in Fig. \ref{fig1} for $r=1/3$ to display its profile in the case the two fields have the same amplitude. The order parameter is now the vector $(\phi,\chi)$ in the plane, which projects as $\phi(x)$ of \eqref{111a} in the $\phi$ axis, and as $\chi(x)$ of \eqref{111b} in the $\chi$ axis. In magnetic materials, the two solutions can be related to the right and left chiralities of the magnetization, since the second field $\chi$ can now be used to control the clockwise or anticlockwise variation of the magnetization with the same energy $E=4/3$. We notice that $r$ controls the width of the Bloch wall and the amplitude of the $\chi$-field solution. The profile in Fig. \ref{fig1} is very much similar to the profile of a Bloch wall that appears in magnetic materials. In this sense, the model provides an interesting representation for the Bloch wall, so we get inspiration from the above analytical calculations to further study modifications that are of current interest in high energy physics, with direct applications in other areas of nonlinear science, in particular in magnetic materials. We further notice that the above model preserves the symmetry of the Bloch wall under the change of chirality and this is of current interest, since chirality may lead to novel effects in some specific magnetic materials: in \cite{CDW1}, in particular, the authors identified a novel chiral Bloch wall, which was shown to appear in Fe/Ni/Cu films; also, in \cite{CDW2} the investigation was able to suggest that the wall chirality could appear induced by Dzyaloshinskii-Moriya interaction \cite{DM1,DM2} in multilayered ferromagnetic elements.

%%%%%%%%%%%%%%%%%%%%
\section{Structure of Bloch Walls}\label{CBW}
 
Inspired by the recent work \cite{multikink}, we consider another model, described by the three real scalar fields, $\phi$, $\chi$ and $\psi$. It is given by the Lagrange density
\begin{equation}\begin{aligned}\label{l1d}
    \mathcal{L}&=\frac{1}{2}f(\psi)\partial_{\mu}\phi\partial^{\mu}\phi+\frac{1}{2}f(\psi)\partial_{\mu}\chi\partial^{\mu}\chi \\
    &\quad+\frac{1}{2}\partial_{\mu}\psi\partial^{\mu}\psi-V(\phi,\chi,\psi).
\end{aligned}
\end{equation}
Here, $f(\psi)$ is a positive function that depends only on the scalar field $\psi$ and the potential is denoted by $V(\phi,\chi,\psi)$. The presence of $f(\psi)$ in this model is intended to modify the mechanical properties associated of the two degrees of freedom $\phi$ and $\chi$. As we will see below, it is capable of simulating geometrical constrictions that change the internal structure of the corresponding solutions. For more information, see Refs. \cite{multikink,BMM,R,L}. To conduct the investigation, since we are dealing with a non canonical model, we follow the lines of Ref.~\cite{genkink2} to get the equations of the general model. The results are given by
\bes\label{eom1d}
\bal
&\partial_{\mu}(f\partial^{\mu}\phi)+V_{\phi}=0,\\
&\partial_{\mu}(f\partial^{\mu}\chi)+V_{\chi}=0,\\
&\partial_{\mu}\partial^{\mu}\psi-\frac{1}{2}f_{\psi}(\partial_{\mu}\phi\partial^{\mu}\phi+\partial_{\mu}\chi\partial^{\mu}\chi)+V_{\psi}=0 ,
\eal
\ees
where $V_{\phi}=\partial V/\partial \phi$, $V_{\chi}=\partial V/\partial \chi$, $V_{\psi}=\partial V/\partial \psi$ and also, $f_{\psi}=d f/d \psi$. Moreover, invariance on spacetime translation gives rise to the energy momentum tensor
\begin{equation}\label{tmunu1d}
    T_{\mu\nu}=f(\psi)\!\left(\partial_{\mu}\phi\partial_{\nu}\phi+\partial_{\mu}\chi\partial_{\nu}\chi\right)+\partial_{\mu}\psi\partial_{\nu}\psi-\eta_{\mu\nu}\mathcal{L}.
 \end{equation}

We consider static configurations. In this case, the equations of motion \eqref{eom1d} take the form
\bes\label{eom1dstatic}
\bal
&(f\phi^\prime)^\prime=V_{\phi},\\
&(f\chi')^\prime=V_{\chi},\\
&\psi''-\frac{1}{2}f_{\psi}\!\left({\phi'}^2+{\chi'}^2\right)=V_{\psi},
\eal
\ees
The non vanishing components of the energy-momentum tensor in Eq.~\eqref{tmunu1d} are the energy density and the stress
\begin{subequations}
	\begin{align}\label{energyd}
\rho &=\frac{1}{2}f(\psi)\left({\phi'}^{2}+{\chi'}^{2}\right)+\frac{1}{2}{\psi'}^{2}+V(\phi,\chi,\psi),\\
p &= \frac{1}{2}f(\psi)\left({\phi'}^{2}+{\chi'}^{2}\right)+\frac{1}{2}{\psi'}^{2}-V(\phi,\chi,\psi).
\end{align}
\end{subequations}
The equations of motion \eqref{eom1dstatic} are of second order with couplings between the fields. To get first order equations, we develop the Bogomol'nyi procedure \cite{multikink,bogo} for this model. To do so, we introduce an auxiliary function $W=W(\phi,\chi,\psi)$ and write the above energy density as
\begin{equation}
\begin{aligned}
\rho &=\frac{f}{2}\left(\phi'\mp\frac{W_{\phi}}{f}\right)^{2}+\frac{f}{2}\left(\chi'\mp\frac{W_{\chi}}{f}\right)^{2}+\frac{1}{2}\left(\psi'\mp W_{\psi}\right)^{2}\\
    &+V -\frac{1}{2}\left(\frac{W^{2}_{\phi}}{f}+\frac{W^{2}_{\chi}}{f}+W^{2}_{\psi}\right) \pm W',
\end{aligned}
\end{equation}
where $W_{\phi}=\partial W/\partial \phi$, $W_{\chi}=\partial W/\partial \chi$, $W_{\psi}=\partial W/\partial \psi$ and $W'$ has the form
\begin{equation}
    W^\prime=\frac{dW}{dx}=W_\phi\;\frac{d\phi}{dx}+W_\chi\;\frac{d\chi}{dx}+W_\psi\;\frac{d\psi}{dx}.
\end{equation}
We now follow the standard procedure and consider the class of potentials that can be written in the form
\begin{equation}
\label{potentialw}
    V(\phi,\chi,\psi)=\frac{1}{2}\left(\frac{W^{2}_{\phi}}{f}+\frac{W^{2}_{\chi}}{f}+W^{2}_{\psi}\right).
\end{equation}
In this situation, the energy density becomes
\begin{equation}
	\begin{aligned}   
    \rho &=\frac{f}{2}\left(\phi'\mp\frac{W_{\phi}}{f}\right)^{2}+\frac{f}{2}\left(\chi'\mp\frac{W_{\chi}}{f}\right)^{2}\\
    &+\frac{1}{2}\left(\psi'\mp W_{\psi}\right)^{2}\pm W'.
\end{aligned}
\end{equation}
The integration of the above expression allows us to see that the energy is bounded,
\be\label{ebogo}
E\geq E_B = \left|\Delta W\right|,
\ee
in which we have $\Delta W = W(\phi(\infty),\chi(\infty),\psi(\infty))-W(\phi(-\infty),\chi(-\infty),\psi(-\infty))$. The minimum energy of the system, $E=E_B$, is attainable when the first order equations below are satisfied
\bes\label{fo1d}
\bal
\label{phi11}
&\phi'=\pm\frac{W_{\phi}}{f},\\
\label{chi11}
&\chi'=\pm\frac{W_{\chi}}{f},\\
\label{psi111}
&\psi'=\pm W_{\psi}.
\eal
\ees
The equations with upper and lower sign are related by the change $x\to-x$. As it was explicitly shown in Ref.~\cite{genkink2}, one can consider a scaling argument in the energy associated to \eqref{energyd} to verify that solutions of the above equations are stable against contractions and dilations, satisfying the stressless condition, $T_{ij}=0$. 

Notice that, since $W$ is, in principle, a function that may present couplings of the three fields, and $f$ depends on $\psi$, the above equations are coupled and must be solved as a system. Nevertheless, if we choose the auxiliary function in the form
\be\label{w12}
W(\phi,\chi,\psi) = W_1(\phi,\chi) + W_2(\psi),
\ee
the first order equation \eqref{psi111} will not depend on the fields $\phi$ and $\chi$, so the field $\psi$ can be solved independently. By knowing the solution $\psi(x)$, one can feed the function $f(\psi)$ in Eqs.~\eqref{phi11} and \eqref{chi11} to find solutions for $\phi(x)$ and $\chi(x)$. In this sense, $\psi$ can be seen as a source for the other fields. Indeed, by using the first order equations \eqref{fo1d}, one can show that the energy density in Eq.~\eqref{energyd} may be written as $\rho=\rho_1+\rho_2$, where
\begin{subequations}
\begin{align}\label{rho11d}
    \rho_1 &= f(\psi)({\phi'}^{2}+{\chi'}^{2}),\\ \label{rho21d}
    \rho_2 &= {\psi'}^{2}.
\end{align}
\end{subequations}
Thus, the contribution of the energy density $\rho_2$, associated to $\psi$, does not depend on both $\phi$ and $\chi$. Note that although the energy density $\rho_1$ depends explicitly on $f$, the same does not happen with the total energy $E_{B}$, which is given in terms of the difference between $W$ calculated at the boundary values of the fields.

%%%%%%%%%%
	\begin{figure}[t!]
		\centering
		\includegraphics[width=7cm,trim={0cm 0cm 0 0},clip]{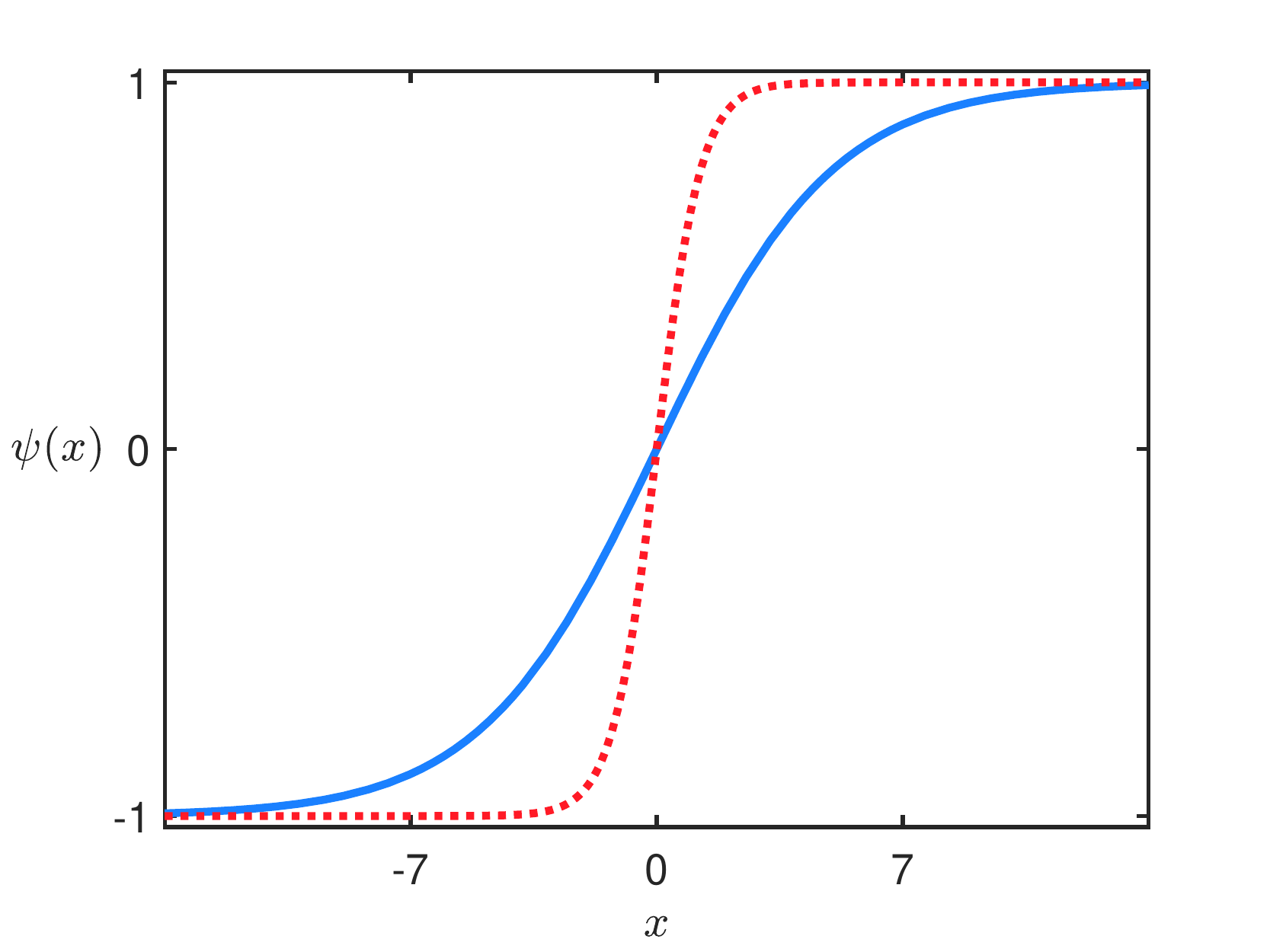}
		\caption{The solution $\psi(x)$ in Eq.~\eqref{solutionpsi}, depicted for $\alpha=0.2$ (solid, blue line) and $\alpha=0.8$ (dotted, red line).}
		\label{fig2}
	\end{figure}
%%%%%%%%%%

The above procedure is important because it shows how we can use the scalar field $\psi$ to modify the behavior of the two other fields $\phi$ and $\chi$. In this paper we want to explore this possibility studying, in particular, how the presence of the function $f(\psi)$ modifies the profile of the fields $\phi$ and $\chi$, which is here used to mimic the so called Bloch wall configuration. To implement the investigation, we get inspiration from Refs.~\cite{multikink} and consider the model described by the auxiliary function in Eq.~\eqref{w12} with
\bes
\bal
\label{W1}
& W_{1}=\phi-\frac{1}{3}\phi^{3}-r\phi\chi^{2},\\
\label{W1b}
& W_{2}=\alpha\psi-\frac{1}{3}\alpha\psi^{3},
\eal
\ees
where $r$ and $\alpha$ are real and positive parameters. In this case, the first order equations \eqref{phi11}, \eqref{chi11} and \eqref{psi111} become
\bes
\bal
\label{FO}
&\phi'=\pm\frac{1-\phi^{2}-r\chi^{2}}{f(\psi)},\\
\label{chisolu}
&\chi'=\mp\frac{2r\phi\chi}{f(\psi)},\\
\label{psisolu}
&\psi'=\pm\alpha(1-\psi^{2}).
\eal
\ees
Notice that Eq.~\eqref{psisolu} depends only on $\psi$ and supports the solution
\begin{align}
    \label{solutionpsi}
    \psi(x)=\tanh{(\alpha x)}.
\end{align}
Using Eq.~\eqref{rho21d} we obtain the energy density 
\begin{align}\label{rho12}
     \rho_{2}=\alpha^{2}\sech^4{(\alpha x)}.
\end{align}
In Fig.~\ref{fig2} we depict the solution \eqref{solutionpsi} for some values of $\alpha$. Integrating $\rho_{2}$ we get $E_{2}= 4\alpha/3$. Notice that the parameter $\alpha$ controls the width and energy of the solution.

In order to determine the solutions $\phi$ and $\chi$, we follow the lines of Ref.~\cite{bflrorbit} and use the trial orbit method to determine the orbit that relates the aforementioned fields. We briefly remind here that the trial orbit method described in \cite{Raja,bflrorbit} consists of suggesting an orbit that relates the fields, as $F(\phi,\chi)=C$, where $F$ is in principle an arbitrary function of the two fields and $C$ is a constant, and verifying if it is compatible with the first order equations \eqref{FO} and \eqref{chisolu} in the topological sector under investigation. In the present case, we consider the elliptic orbit $a\phi^2+b\chi^2=1$, with $a$ and $b$ constants, and the topological sector defined by two minima $v_{1}=(-1,0)$ and $v_{2}=(1,0)$, for the pair of fields $(\phi,\chi)$. The restriction that the orbit has to contains the points $(1,0)$ and $(-1,0)$ leads to $a=1$. This implies that $\phi\phi^{'} +b \chi\chi^{'}=0$, and using Eqs.~\eqref{FO} and \eqref{chisolu} we get that $b=r/(1-2r)$, for $r\in (0,1/2)$. Thus, the elliptic orbit
\begin{equation}\label{orbit}
    \phi^{2}+\frac{r}{1-2r}\chi^{2}=1,
\end{equation}
is supported by our model. It is interesting to see that the above orbit does not depend on $f(\psi)$, and this will help us to find explicit solutions in this new model. In particular, we notice that even though the field $\psi$ is explicitly determined, the form of the potential in Eq.~\eqref{potentialw} and the fields $\phi$ and $\chi$ obtained via the first order equations \eqref{FO} and \eqref{chisolu} depend on the function $f(\psi)$. However, the orbit in Eq.~\eqref{orbit} is valid for any positive function $f(\psi)$ and can be used to decouple the first order equation \eqref{FO}, which now reads
\be\label{fophiorbit}
\phi'=\pm\frac{2r(1-\phi^{2})}{f(\psi)}.
\ee

This equation is quite similar to the one which arises in Refs.~\cite{BMM,multikink}, where geometrically constrained kinks were investigated. Since it depends explicitly on the choice of $f(\psi)$, we may consider this function to modify the core of the solutions, giving rise to the presence of internal structure. Furthermore, since $\psi=\psi(x)$, the function $f(\psi)$ will be, ultimately, a function depending on the spatial coordinate. In this sense, one may wonder if considering a spatially dependent $f(x)$ instead of $f(\psi)$, without adding a third field, would suffice. However, the presence of  $f(x)$ directly in the Lagrange density \eqref{l1d} would break the translational invariance of the model. Thus, the presence of the third field, $\psi$, allows us to modify the solutions $\phi$ and $\chi$ preserving the translational invariance of the model and keeping the stability of the solutions. Next, we investigate distinct possibilities.

\subsection{First Model}\label{model1da}

\begin{figure}[t!]
		\centering
		\includegraphics[width=7cm,trim={0cm 0cm 0 0},clip]{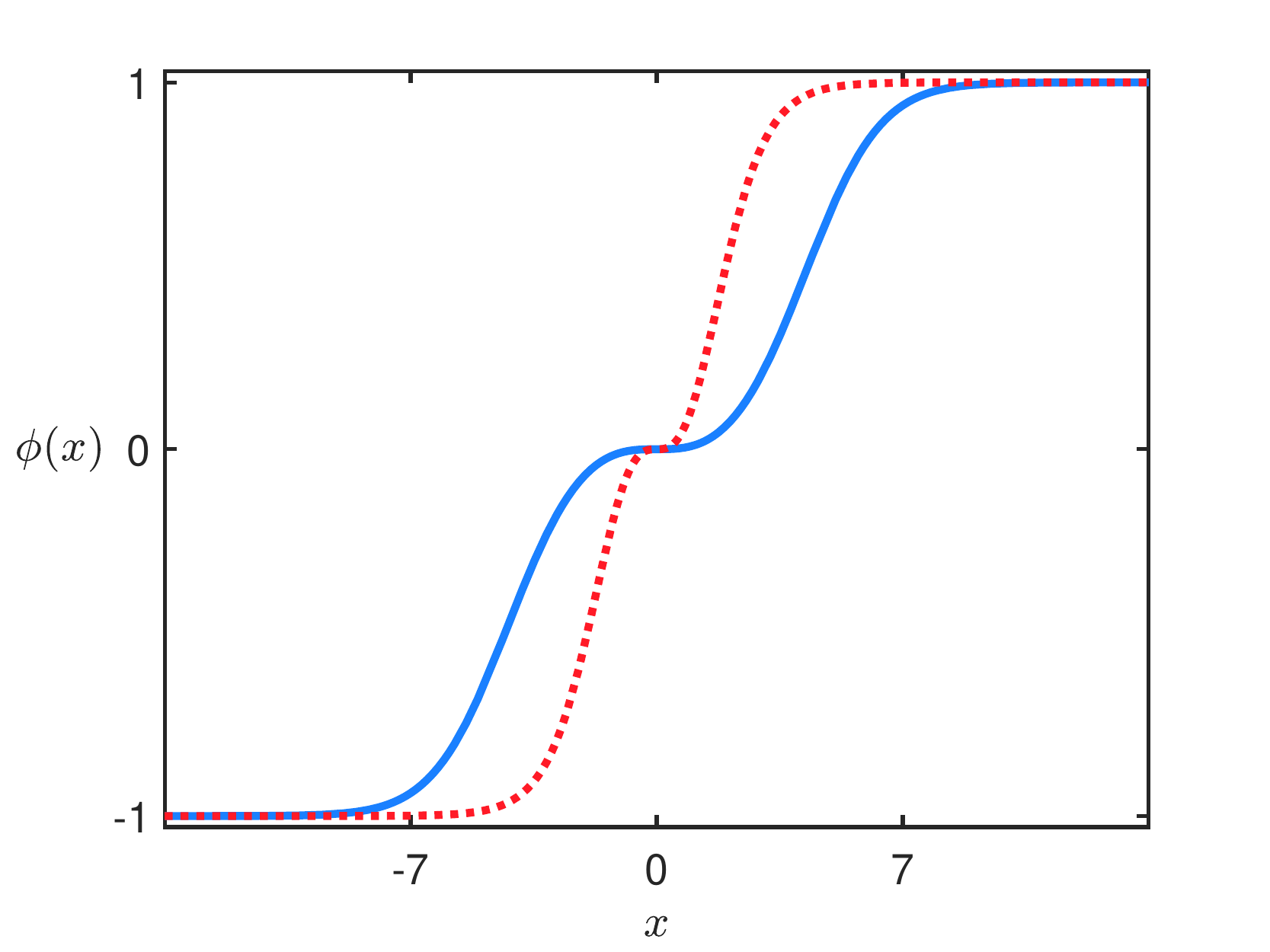}
		\includegraphics[width=7cm,trim={0cm 0cm 0 0},clip]{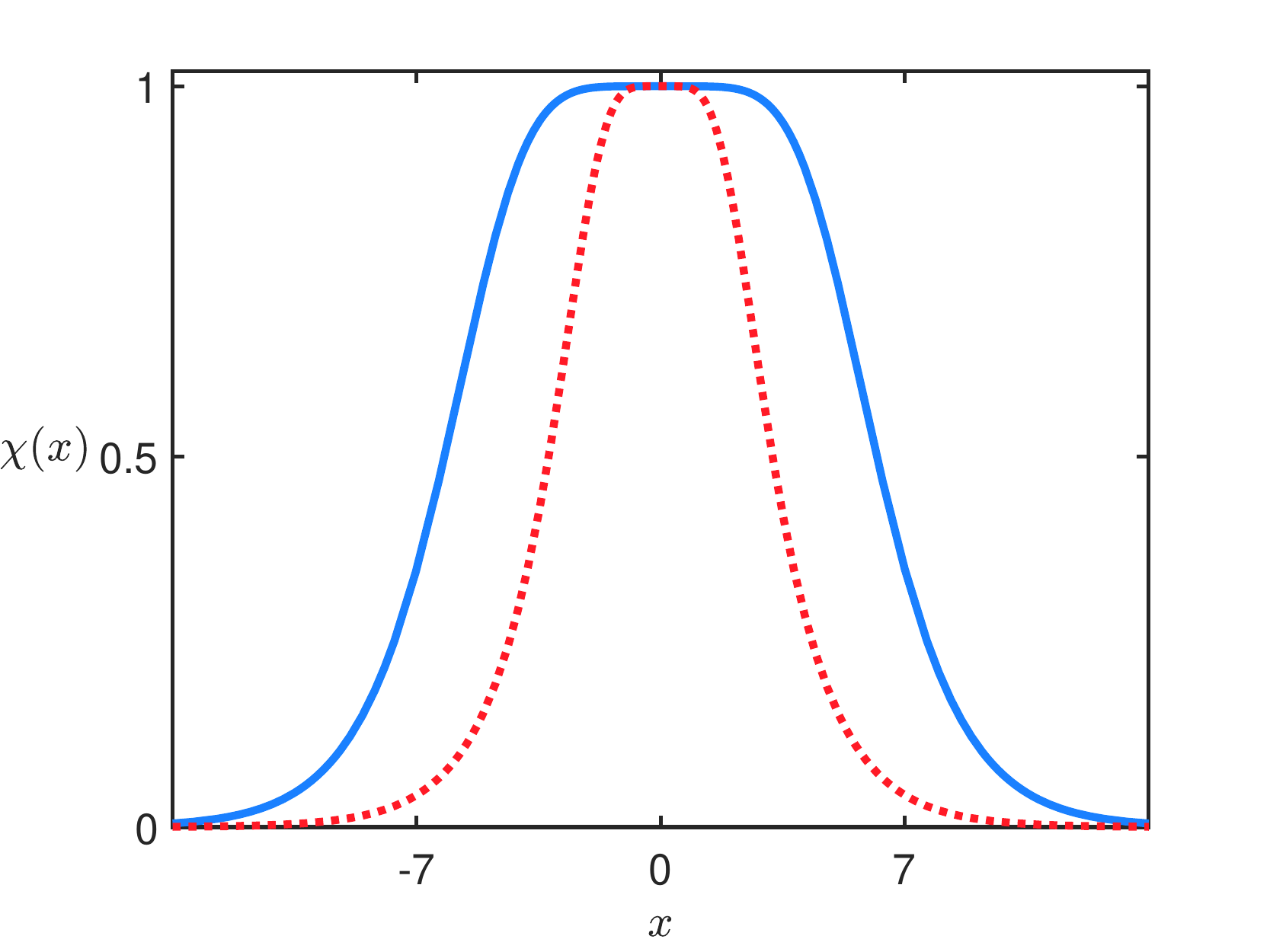}
		\caption{The solutions $\phi(x)$ (top) and $\chi(x)$ (bottom) associated to the model in Sec.~\ref{model1da}, for $r=1/3$ and $\alpha=0.2$ (solid, blue line), and $0.8$ (dotted, red line).}
		\label{fig3}
		\end{figure}

The first model arises with $f(\psi)=1/\psi^{2}$. In this case, the potential in Eq.~\eqref{potentialw} becomes 
\begin{equation}
	\begin{aligned}
    V(\phi,\chi,\psi)&=\frac{1}{2}\psi^{2}(1-\phi^{2}-r\chi^{2})^{2}\\
    &+2r^2\phi^2\chi^2\psi^2+\frac12\alpha^{2}(1-\psi^{2})^{2}.
\end{aligned}
\end{equation}
This choice of $f(\psi)$ is inspired by Ref. \cite{R}, as we further comment on below. To determine the solutions $\phi$ and $\chi$, one must feed the function $f(\psi)$ with the solution \eqref{solutionpsi} in the first order equations \eqref{FO} and \eqref{chisolu}. Since $\psi(x)$ is given by Eq.~\eqref{solutionpsi}, we have $f(\psi(x)) = \tanh^{-2}{(\alpha x)}$, which diverges at $x=0$, and the first order equation \eqref{fophiorbit} takes the form
\begin{equation}
\label{firstoeq}
    \phi'=2r(1-\phi^{2})\tanh^{2}{(\alpha x)}.
\end{equation}
Notice that the divergence which arises in the function $f$ at $x=0$ makes the derivative of $\phi$ vanish at $x=0$, thus adding a plateau in the solution $\phi(x)$ at the origin; see Fig. \ref{fig3}. This is similar to the case of magnetic materials in the presence of geometrical constrictions investigated experimentally in Ref.~\cite{R}, which also induces a plateau in the magnetization of the magnetic material there considered. The above equation admits the analytical solution
\begin{align}
\label{phi1eq}
    \phi(x)=\pm\tanh{(\xi(x))},
\end{align}
where
\begin{equation}\label{xi1d}
\xi(x)=2r(x-\tanh{(\alpha x)}/\alpha).
\end{equation}
Using the above solution and the orbit \eqref{orbit}, we get
\begin{align}
\label{chi1eq}
    \chi(x)=\pm\sqrt{\frac{1}{r}-2}\,\sech{(\xi(x))}.
\end{align}
Since the strength of the coupling between the fields $\phi$ and $\chi$ is controlled by the parameter $r$ in Eq.~\eqref{W1}, let us investigate how it modifies the localized structure. In Fig.~\ref{fig3} we depict the solutions \eqref{phi1eq} and \eqref{chi1eq} for  $r=1/3$ and for $\alpha=0.2$ and $0.8$. We see that $\alpha$ controls the width of the solutions, so it can be used to determine the size of the localized wall. Notice that $\phi$ has a double kink profile and $\chi$ has a plateau at its center. 
%%%%%%%%%%%%%%%%%%%%%%%%%%%
%%%%%%%%%%%%%%%%%%%%%%%%%%%

The energy density associated to $\phi$ and $\chi$ in Eq.~\eqref{rho11} for this model is
\begin{equation}
	\begin{aligned}
    \label{rho11}
\rho_{1}&=4r^{2}\tanh^{2}{(\alpha x)}\,\sech^{2}{(\xi)}\bigg[\sech^{2}{(\xi)}\\
 		&\quad+\tanh^{2}{(\xi)}\bigg(\frac{1}{r}-2\bigg)\bigg].
\end{aligned}
\end{equation}

By integrating it we get $E_{1}=4/3$. The total energy is then $E=E_1+E_2 = 4(1+\alpha)/3$, as expected from Eq.~\eqref{ebogo}. This shows that the energy does not depend on $r$, which is the parameter that controls the coupling of $\phi$ and $\chi$ in the the model. 

\subsection{Second Model}\label{model1db}

We now consider $f(\psi)=\sec^{2}(n\pi\psi)$, where $n$ is an integer. This model is a generalization of the previous one, inspired by Ref.~\cite{BMM,R}, in order to mimic geometrical constrictions at several distinct positions inside  the solution, due to the periodic profile of this new $f(\psi)$ and the integer $n$. This can be of interest to control the domain wall polarity under the presence of fermions, as suggested in Refs. \cite{BMM,L}. The $\phi$ field is governed by Eq.~\eqref{fophiorbit}, which reads
\begin{equation}
    \phi'=2r(1-\phi^{2})\cos^{2}{(n\pi\tanh{(\alpha x)})}.
\end{equation}
From the above equation and the orbit equation \eqref{orbit}, one can show that this model supports the very same solutions in Eqs.~\eqref{phi1eq} and \eqref{chi1eq} with the geometrical coordinate $\xi(x)$ now replaced by $\eta(x)$, such that
\begin{align}\label{geom21d}
    \eta(x)=rx+\frac{r}{2\alpha }\big(\textrm{Ci}(\xi_{+}(x))-\textrm{Ci}(\xi_{-}(x))\big),
\end{align}
where $\xi_{\pm}(x)=2n\pi(1\pm\tanh{(\alpha x)})$.
%%%%%%%%%%%%%%%%%%%%%%%%%%%
		\begin{figure}[t!]
		\centering
		\includegraphics[width=6.8cm,trim={0cm 0cm 0 0},clip]{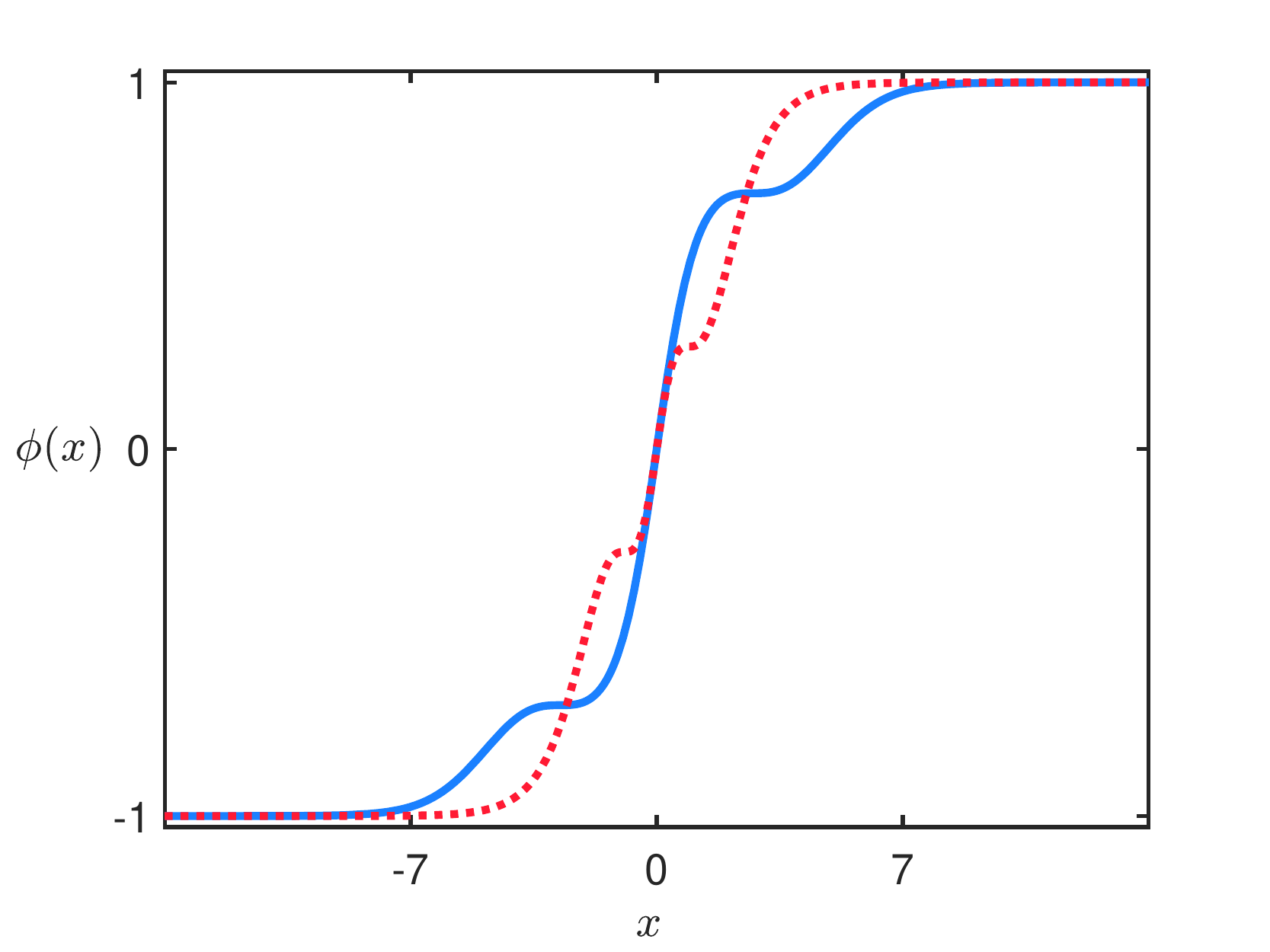}
		\includegraphics[width=6.8cm,trim={0cm 0cm 0 0},clip]{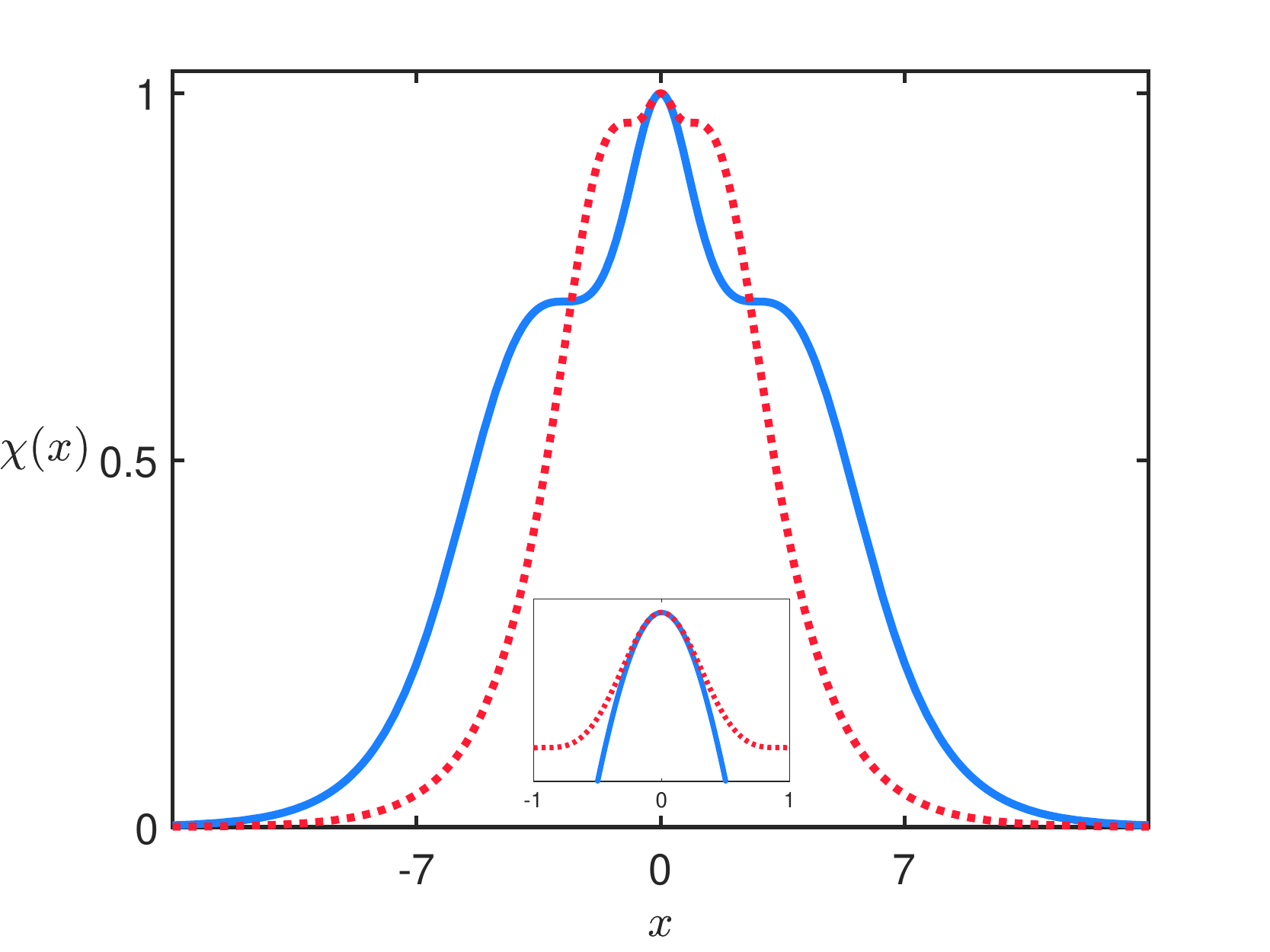}
		\includegraphics[width=6.8cm,trim={0cm 0cm 0 0},clip]{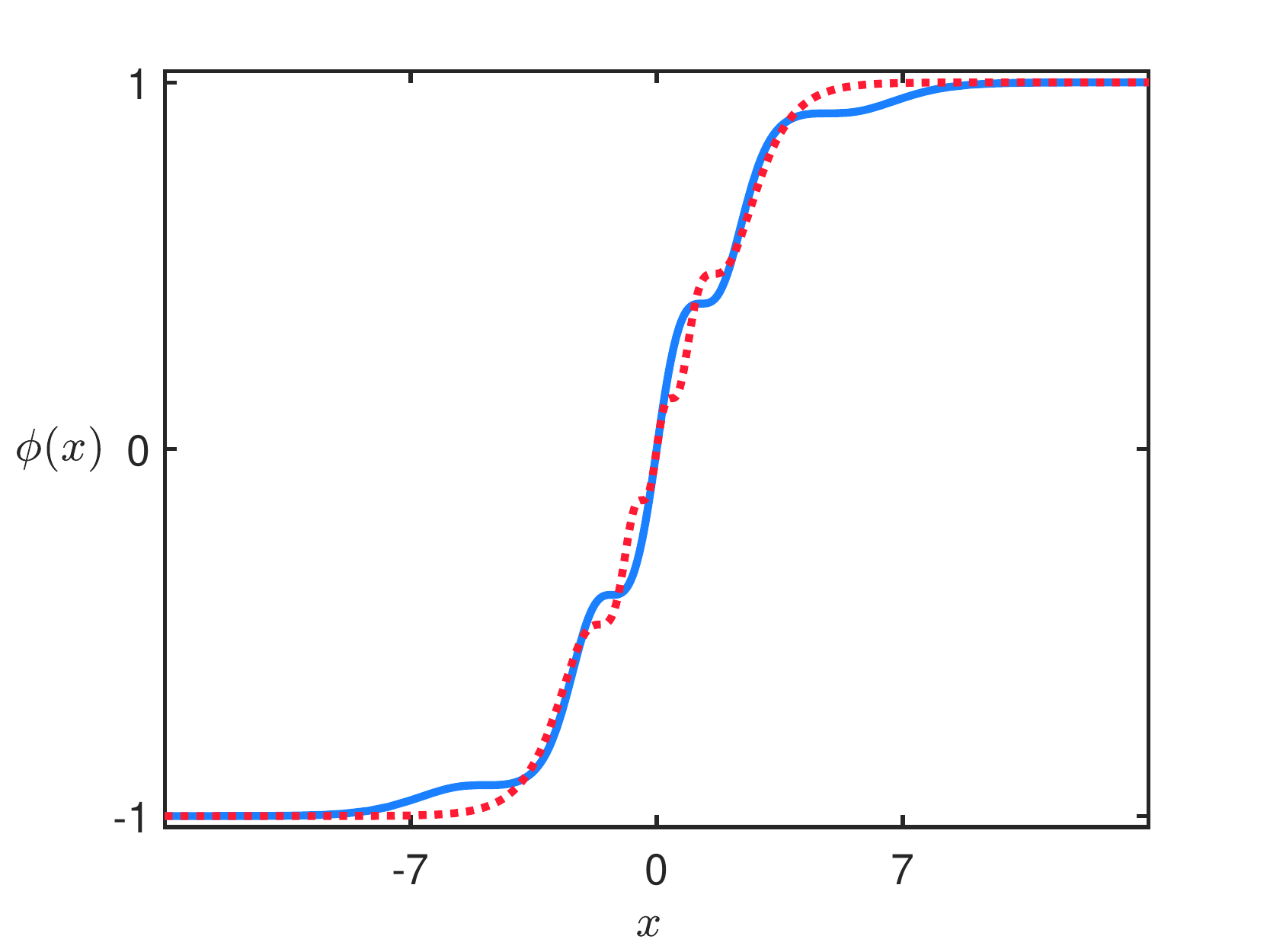}
		\includegraphics[width=6.8cm,trim={0cm 0cm 0 0},clip]{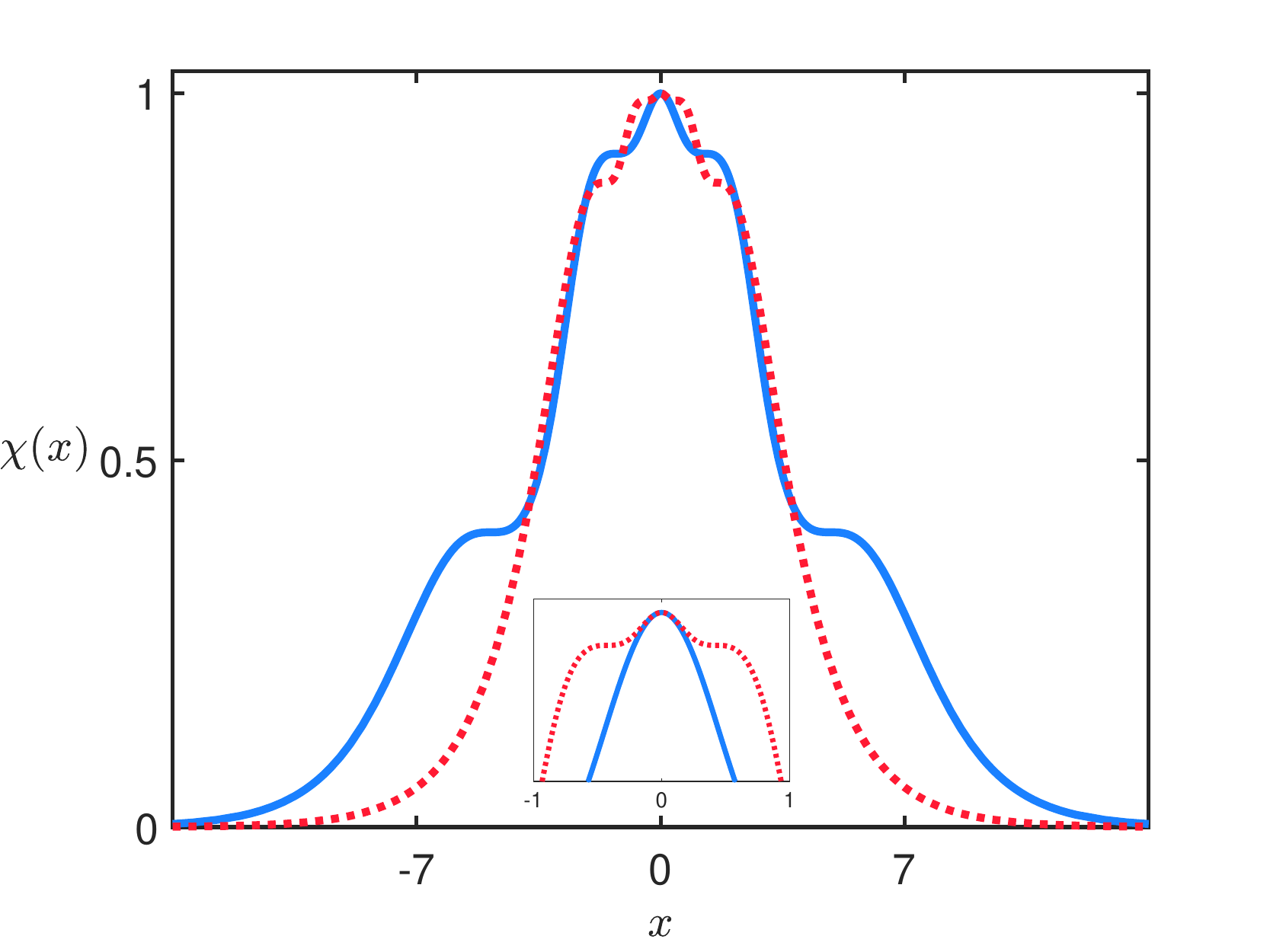}
		\caption{The solutions $\phi(x)$ and $\chi(x)$ for the model in Sec.~\ref{model1db}, with $f(\psi)=\sec^2(n\pi\psi)$, depicted with $r=1/3$ and for $\alpha=0.2$ (solid, blue line) and $0.6$ (dotted, red line) and with $n=1$ (top) and $n=2$ (bottom). The insets highlight the behavior of the $\chi$-field configurations near the origin.}
		\label{fig4}
		\end{figure}
%%%%%%%%%%%%%%%%%%%%%%%%%%%
%%%%%%%%%%%%%%%%%%%%%%%%%%%
		\begin{figure}[p]
		\centering
		\includegraphics[width=6.8cm,trim={0cm 0cm 0 0},clip]{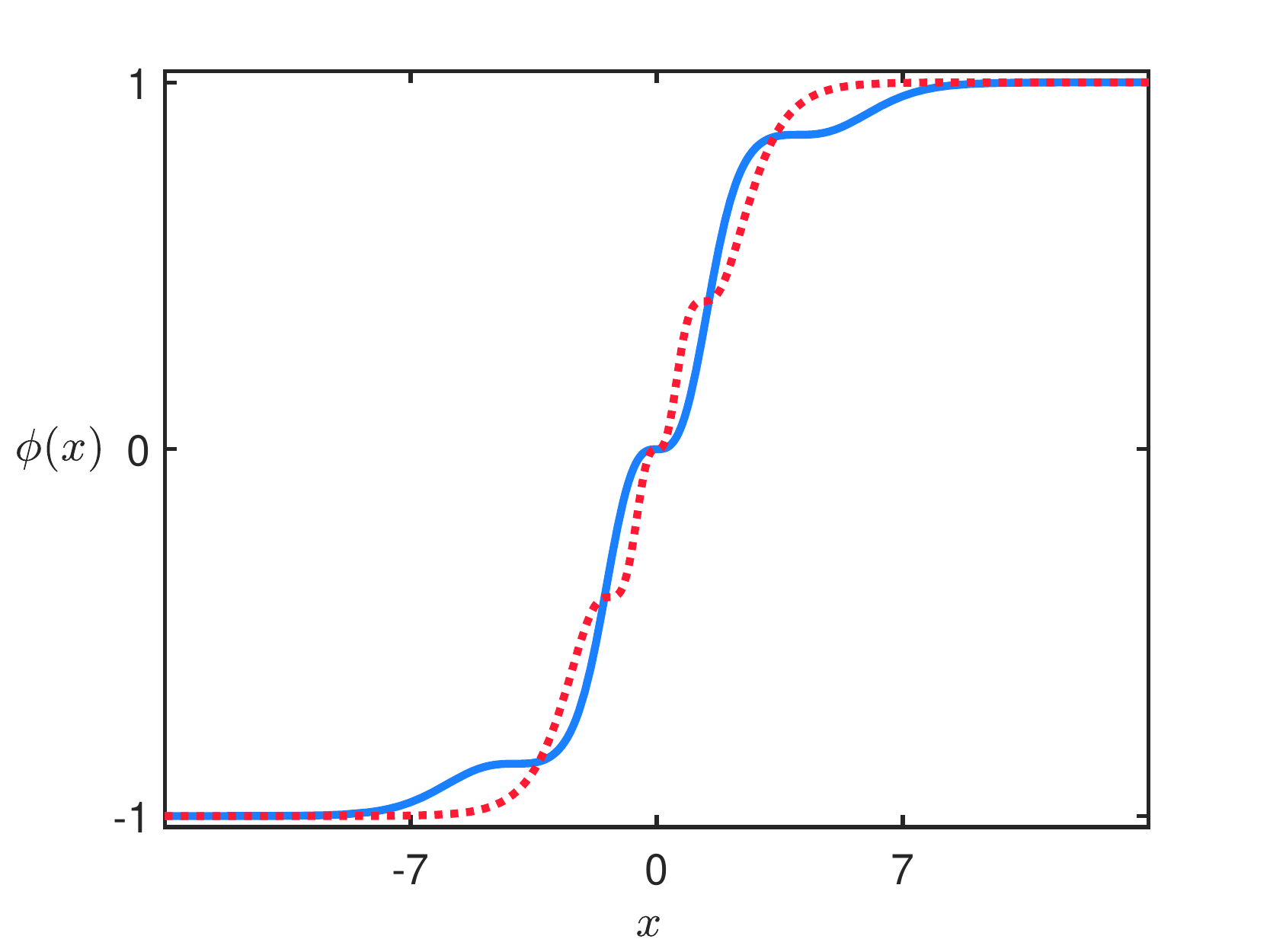}
		\includegraphics[width=6.8cm,trim={0cm 0cm 0 0},clip]{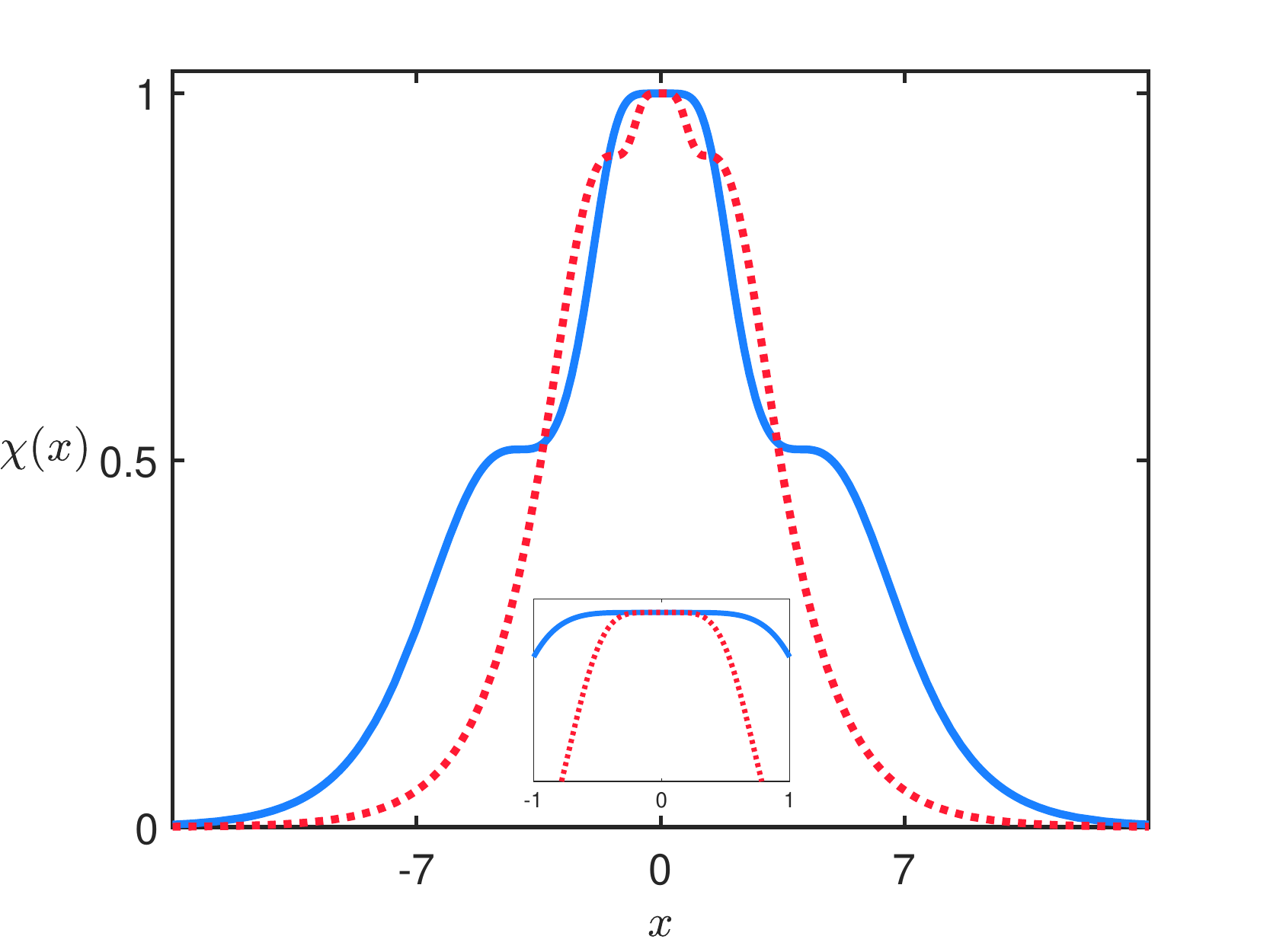}
		\includegraphics[width=6.8cm,trim={0cm 0cm 0 0},clip]{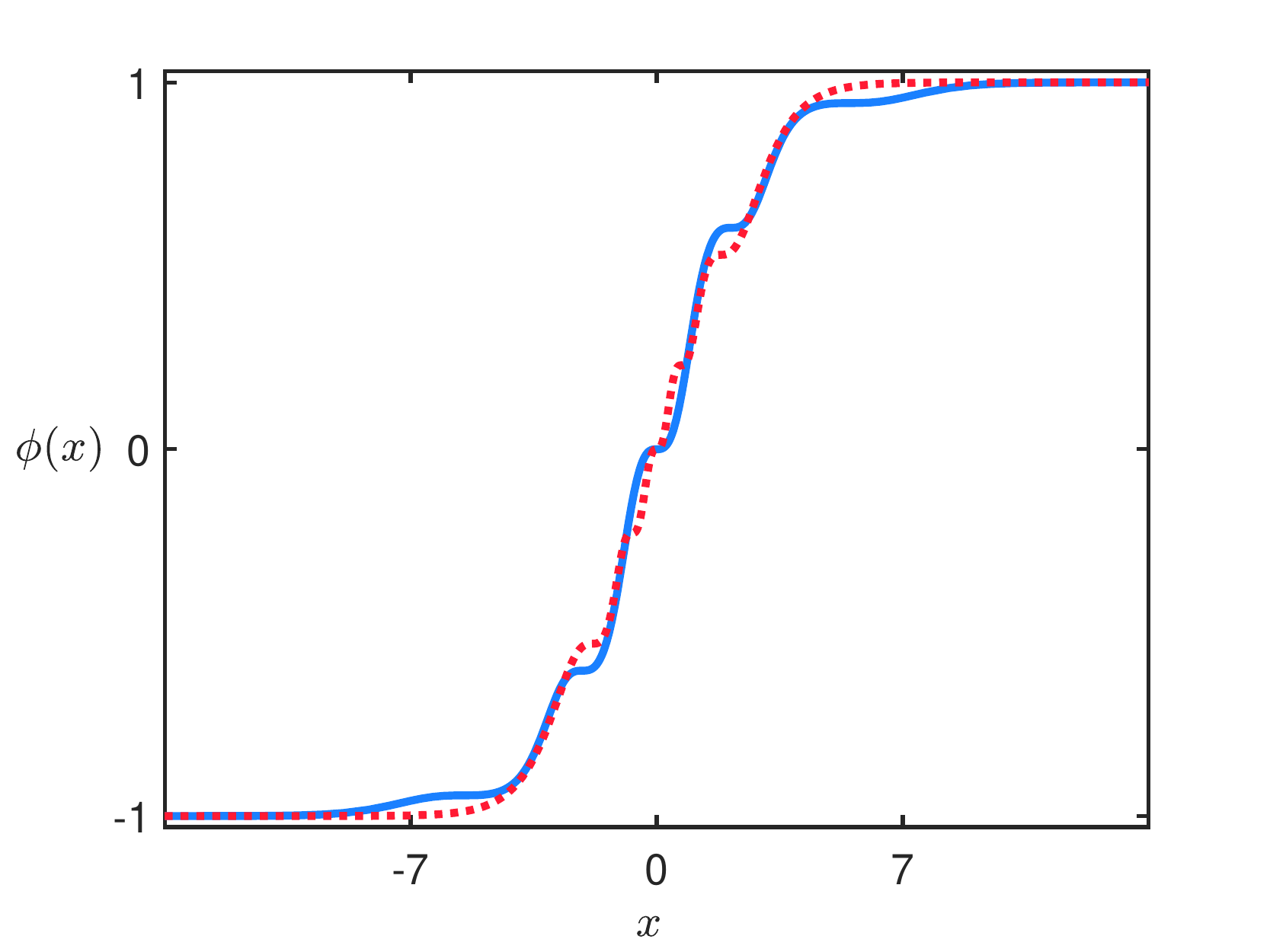}
		\includegraphics[width=6.8cm,trim={0cm 0cm 0 0},clip]{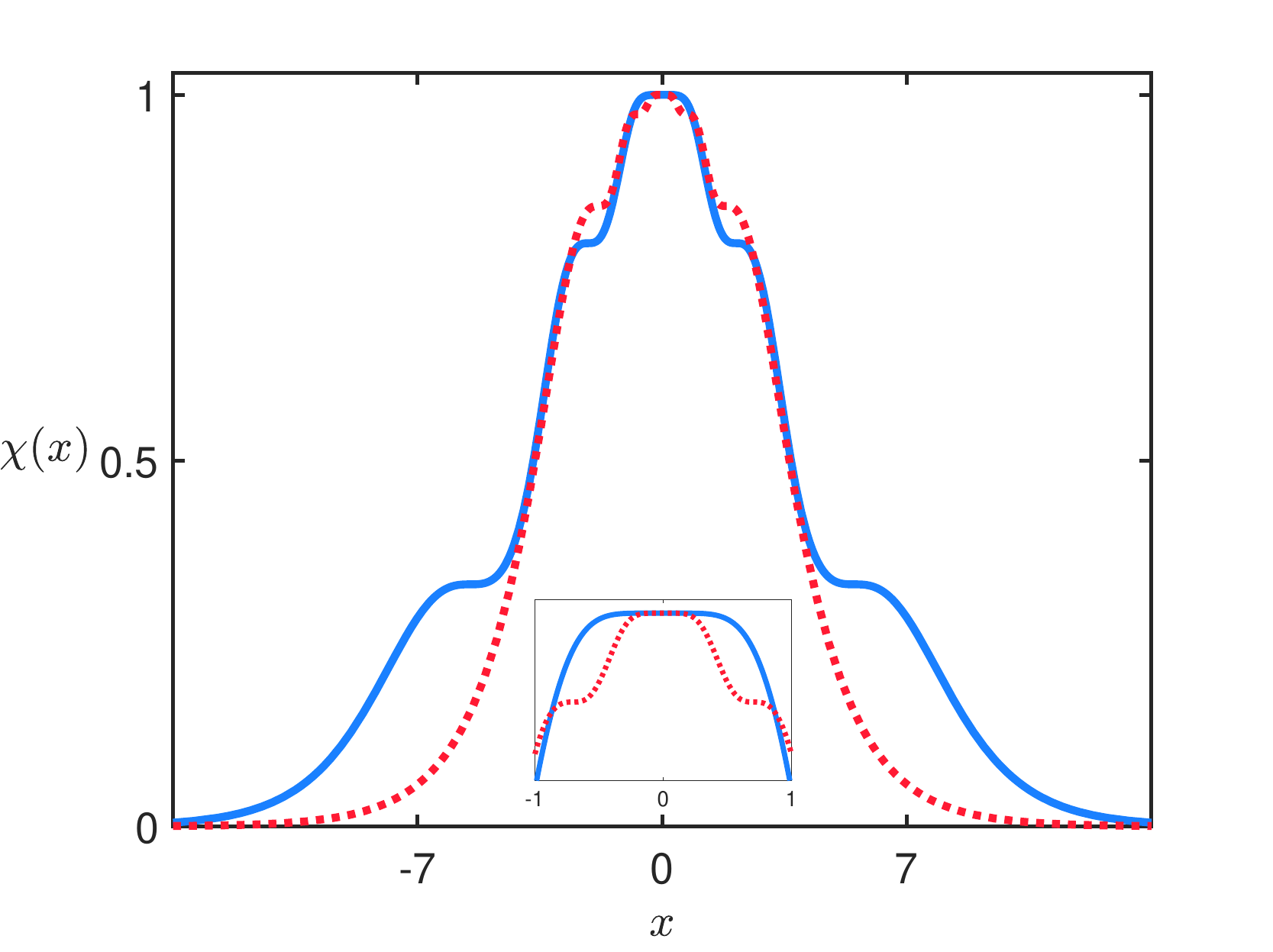}
		\caption{The solutions $\phi(x)$ and $\chi(x)$ for the model in Sec.~\ref{model1db}, with $f(\psi)=\csc^2((n+1/2)\pi\psi)$, depicted with the same values used in Fig. \ref{fig4}, for comparison.  The insets highlight the behavior of the $\chi$-field configurations near the origin.}
		\label{fig5}
		\end{figure}
%%%%%%%%%%%%%%%%%%%%%%%%%%%

In the above expression $\textrm{Ci}(z)$ is defined as
\begin{align}
    \textrm{Ci}(z)=\gamma+\ln(z)+\int^{z}_{0}\frac{\cos(y)-1}{y}dy,
\end{align}
where $\gamma$ is the Euler–Mascheroni constant. For small and large $z$, this function behaves as, respectively,
\bes
\bal
    &\textrm{Ci}(z)=\gamma+\ln{(z)}-\frac{z^{2}}{4}+\mathcal{O}(z^{4}),\\
     &\textrm{Ci}(z)=\frac{\sin{(z)}}{z}-\frac{\cos{(z)}}{z^{2}}+\mathcal{O}(1/z^{3}).
\eal
\ees
The solutions $\phi(x)$ and $\chi(x)$ are displayed in Fig.~\ref{fig4} for $r=1/3$, $\alpha=0.2$ and for some values of $n$. We see that both the kinklike solution $\phi$ and the lumplike solution $\chi$ presents $2n$ inflection points or plateaux, so $n$ may be used to control the internal structure of the wall. As far as we know, this is the first time we have seen this kind of profile in bell-shaped lumplike configurations. It is a direct consequence of the function $f(\psi)$, which now has divergences that can be  controlled by $n$. These divergences induce zeroes in the derivative of the solutions, which lead to the several plateaux we see in Figs.~\ref{fig4} and \ref{fig5}, mimicking the effects of geometrical constrictions in the system, as considered in Refs. \cite{multikink,BMM,R}, for instance.

The energy density \eqref{rho11d} for this model is given by
\begin{align}
    \label{densityed11}
    \rho_{1}&=4r^{2}\cos^{2}{(n\pi\tanh{(\alpha x)})}\sech^{2}{(\eta)}\bigg[\sech^{2}{(\eta)}\nonumber\\
     &\quad+\tanh^{2}{(\eta)}\bigg(\frac{1}{r}-2\bigg)\bigg].
\end{align}
The total energy of the model is the same of the previous one, $E= 4(1+\alpha)/3$, as expected from Eq.~\eqref{ebogo}, since it does not depend on the function $f(\psi)$.

Similarly, we can consider a slightly different function, using $f(\psi)=\csc^{2}((n+1/2)\pi\psi)$. In this case we get

\begin{equation}
    \phi'=2r(1-\phi^{2})\sin^{2}{((n+1/2)\pi\tanh{(\alpha x)})}.
\end{equation}
The solution follows as in the previous case. Here, however, we have to change $\xi_{\pm}(x)=2n\pi(1\pm\tanh{(\alpha x)})$ to the new $\xi_{\pm}(x)=(2n+1)\pi(1\pm\tanh{(\alpha x)})$, and now the number of inflection points in the kinklike configuration $\phi$ is $2n+1$, with the appearance of an extra inflection point at the origin. In Fig. \ref{fig5} we depict the solutions for the same values used in Fig. \ref{fig4}, to better compare with the previous case. This modification changes significantly the profile of the bell-shaped component of the solution, as we see when we compare the two Figs. \ref{fig4} and \ref{fig5}.
%%%%%%%%%%%%%%%%%%%%%%%%%%%
		\begin{figure}[t!]
		\centering
		\includegraphics[width=6.8cm,trim={0cm 0cm 0 0},clip]{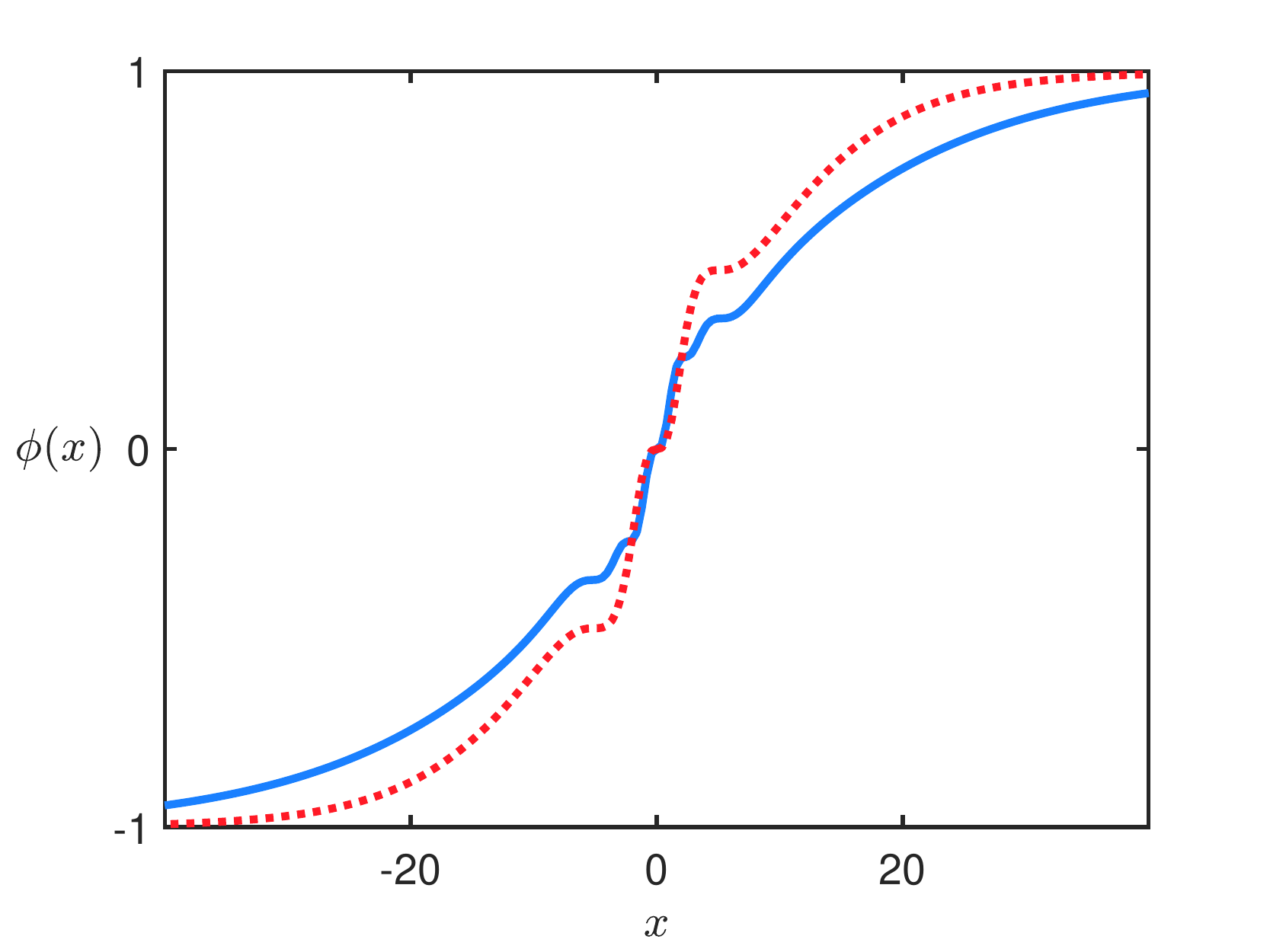}
		\includegraphics[width=6.8cm,trim={0cm 0cm 0 0},clip]{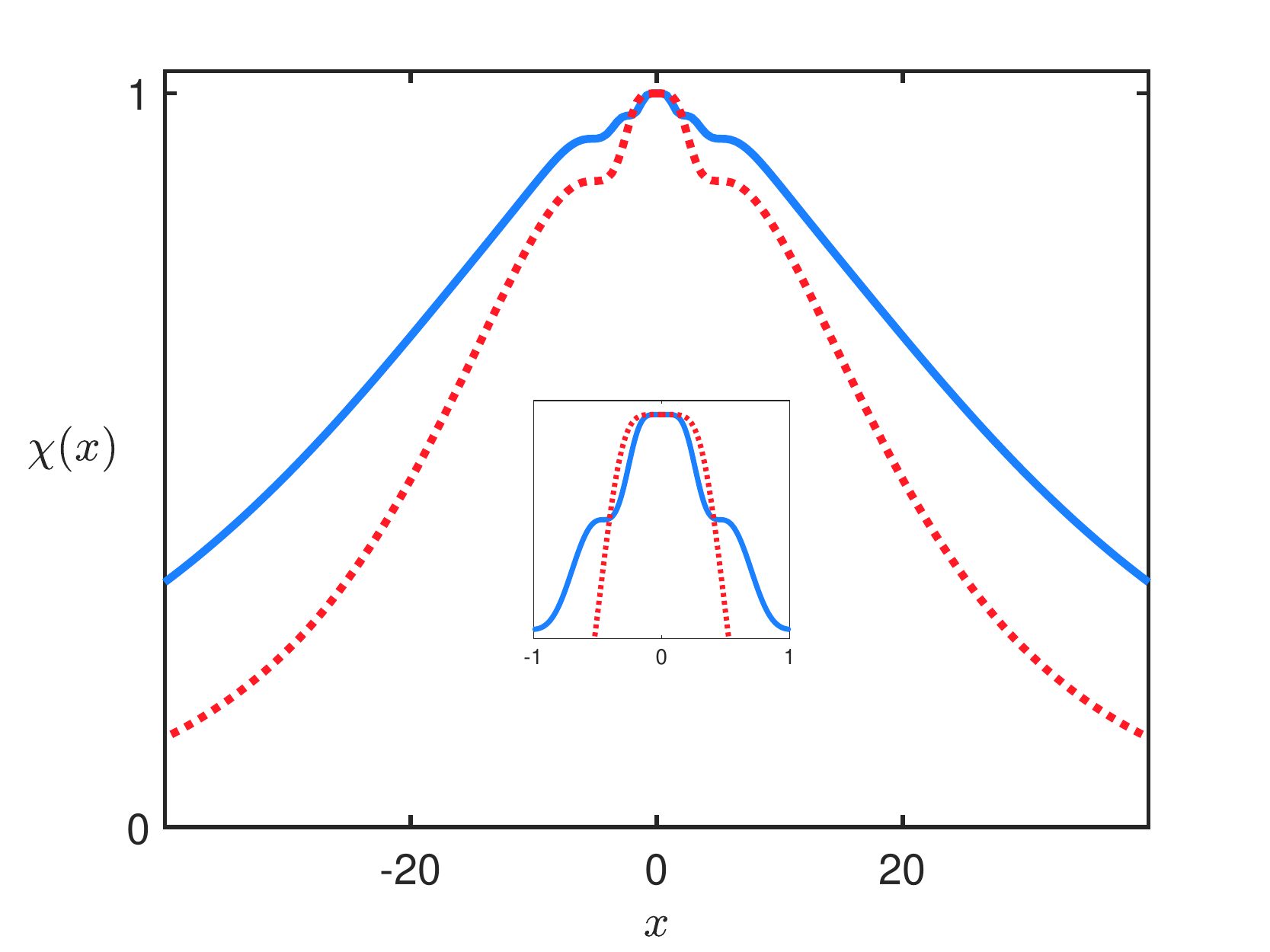}
		\caption{The solutions $\phi(x)$ (top panel) and $\chi(x)$ (bottom panel) for the model in Sec.~\ref{model3da}, with $f(\psi)=1/J^2_1(a\psi)$, depicted with $r=1/3$ and for $\alpha=0.2$ and with $a=5$ (red, dotted line) and $a=9$ (blue, solid line). The inset highlight the behavior of the $\chi$-field configurations near the origin.}
		\label{fig6}
		\end{figure}
%%%%%%%%%%%%%%%%%%%%%%%%%%%

%%%%%%%%%%%%%%%%
\subsection{Third Model}\label{model3da}

We can consider other possibilities for the function $f(\psi)$, for example, the case $f(\psi)=1/J^2_1(a\psi)$, where $J_1$ stands for the Bessel function of first kind and $a$ is a real and positive parameter. This is inspired by Refs. \cite{PRE,PRL1,PRL2,OL,OE}, where the authors deal with the presence of vortices in Bessel optical lattices. Here, we bring this idea to the case of a single spatial dimension, that is, to modulate the one-dimensional medium with the Bessel function of first kind, for instance, to see how it works in the case of Bloch wall. It is of interest to emphasize that the stability of the solution is ensured by the procedure, which works for solutions of first order equations that minimize the total energy of the system. 

With  this choice, the first order equation for $\phi$ becomes
\begin{equation}
    \phi'=2r(1-\phi^{2}) J_1^2(a\tanh(\alpha x)).
\end{equation}
We have solved this equation numerically, and used it to solve for $\chi$ via the orbit constraint displayed in Eq. \eqref{orbit}. Some solutions are depicted in Fig. \ref{fig6}. The profile is somehow similar to the case depicted in Fig. \ref{fig5}, but here the localized structures are larger than they appear in Figs. \ref{fig3}, \ref{fig4} and \ref{fig5}. 

\begin{figure}[t!]
		\centering
		\includegraphics[width=6.8cm,trim={0cm 0cm 0 0},clip]{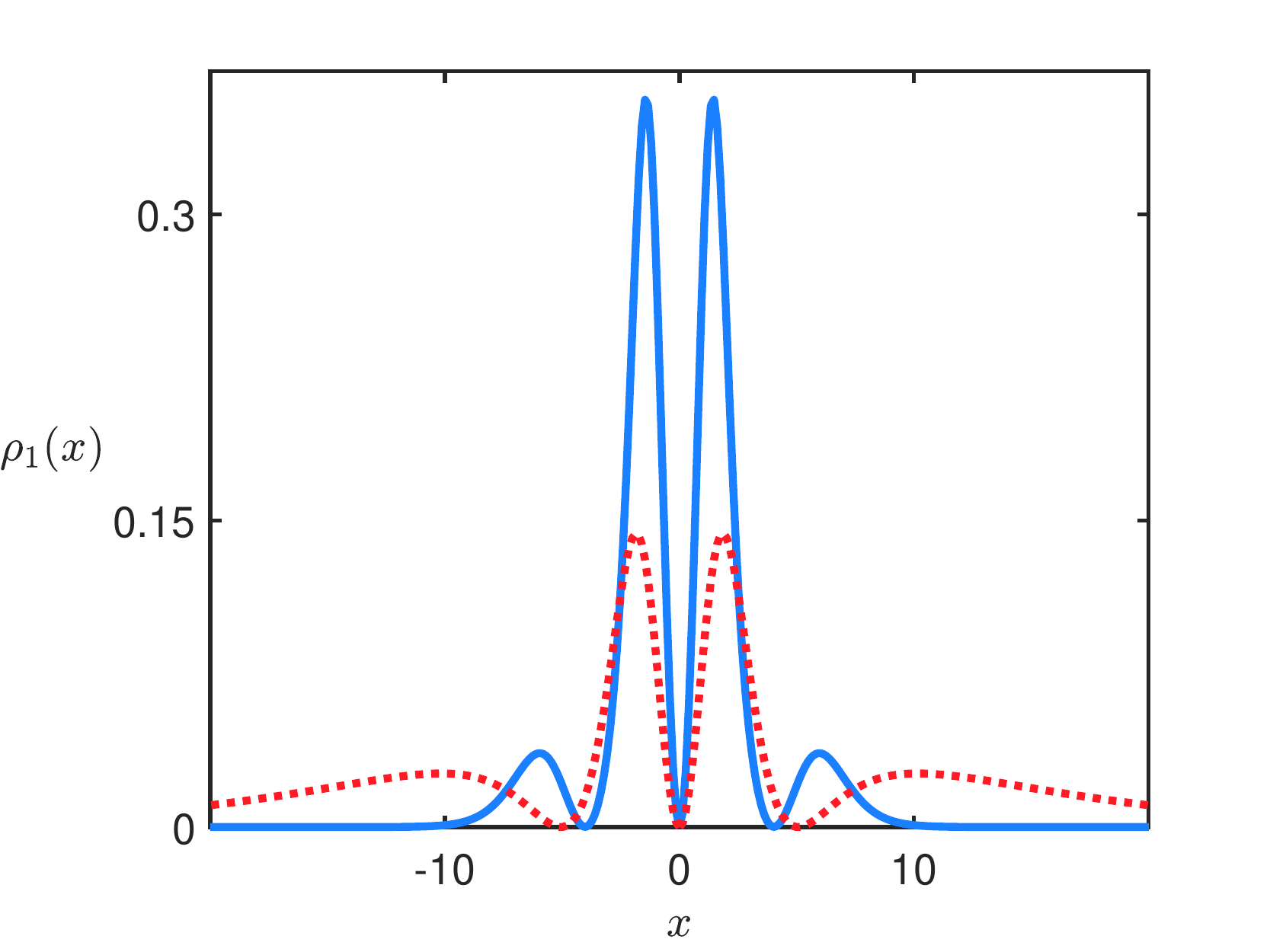}
		\caption{The energy density for $f(\psi)$ being controlled by the cosecant (blue) and the Bessel (red) functions, depicted for $r=1/3$, $\alpha=0.2$, and for $n=1$ and $a=5$, respectively.}
		\label{fig7}
		\end{figure}

As we have seen from the procedure developed above, the energies of the solutions do not depend of the function $f(\chi)$, although it modifies the internal structure of the corresponding configurations. Moreover, we notice that the solutions displayed in Figs. \ref{fig2}, \ref{fig3} and \ref{fig4} and \ref{fig5}, they all have similar size in the range of parameters used to depict them. However, the solutions of the last model that appear in Fig. \ref{fig6} are larger than all the other, so the use of the Bessel function also contributes to enlarge the related domain wall. To make this point clearer, in Fig. \ref{fig7} we depict the energy density for $f(\psi)$ controlled by the cosecant and the Bessel functions, for appropriate choices of the parameters of the two models. Although they have similar profiles, one notices that the energy density corresponding to the Bessel function is much less concentrated around its center.

%%%%%%%%%%%%%%%%%%
%%%%%%%%%%%%%%%%%%%%%%%%%%%
%%%%%%%%%%%%%%%%%%%%%%%%%%%

%%%%%%%%%%%%%%%%%%%%%%%%%%%%
\section{Conclusions}\label{conclu}

In this work, we have investigated a three-field model, in which a single scalar field is included to modify the kinematics of a two-field configuration that behaves as a Bloch wall. We have developed a first order formalism based on the energy minimization of the system. In this situation, we have found a way to decouple the additional field from the other ones, which may be used to modify the geometry of the Bloch wall. We have illustrated our procedure with several distinct models, most of them presenting analytical solutions, and the last one which is solved numerically. The models are controlled by distinct functions of the third field, $f(\psi)$, which are used to modify the medium where the two other fields interact, and this changes significantly the internal structure of the corresponding Bloch walls, so they can be considered in applications of practical interest. 

As we have shown, the internal plateaux which are added to the standard Bloch wall in the presence of the third field is similar to the effects of geometric constriction investigated before in Ref. \cite{R}, which directly contributed to induce a plateau at the center of the hyperbolic tangent profile associated to the magnetization of the magnetic Fe$_{20}$Ni$_{80}$ element considered in the corresponding experiment. The results of the present work are robust and provide an interesting theoretical framework that allows the search for generalized solutions that are stable, indicating that Bloch walls subjected to geometric constrictions at the nanometric scale may change the internal disposition of magnetization. This is an effect of current interest for the manipulation of the magnetic information at small distances, not present in the conventional Bloch wall configuration.

The study suggests that we further investigate other possibilities, modifying the potential that controls the fields $\phi$ and $\chi$, and the function $f(\psi)$. We can, for instance, suppose that $\phi$ and $\chi$ are described by non-polynomial functions of the trigonometric or hyperbolic type. As one knows, the description of Kerr media in optical fibers \cite{OS} and negative capacitance in ferroelectric systems \cite{JAP} are usually considered with the help of polynomial interactions; however, in magnetic systems one may find useful to describe the magnetization vector with trigonometric functions \cite{PhysRevB} and this motivates us to further consider trigonometric interactions. The use of hyperbolic functions may also be useful, since it may give rise to models engendering kink profile, and may describe specific behavior; see, e.g., the investigations on the scattering of kinks in models with hyperbolic \cite{Gani,Gomes} and double sine-Gordon \cite{Azadeh} interactions. Polynomial potentials with higher order power in the scalar field are also of current interest, as recently studied in \cite{Gani2} for kink-kink and kink-antikink collisions. We can also consider the collision with an impurity, which, under certain circumstances, may result in a backward scattering \cite{impu}. Another issue of current interest refers to the addition of fermions, to see how the fermionic degrees of freedom behave under the presence of the modified Bloch walls. This last possibility is similar to the case recently studied in \cite{BMM}, and is presently under further consideration. 

In condensed matter, the study of the Bloch wall chirality is also of current interest, in connection with the effects unveiled in Refs. \cite{CDW1,CDW2}. In particular, we can consider the addition of new contributions in the models explored in the present work, searching for the breaking of chirality, to relate this with the investigation of \cite{CDW2}, where an interlayer Dzyaloshinkii-Moriya interaction \cite{DM1,DM2} was used to break the degeneracy between Bloch wall chiralities. We can also use of the Bloch walls investigated above to study composite domain walls in multiferroic materials \cite{N1,N2}; see, e.g., \cite{RevN0,RevN,RevN2} for further issues on this topic. In particular, trigonometric interactions have been recently considered in \cite{RevN2} for the interconversion of domain to domain walls and this also suggests other possibilities directly related to the topic of Bloch wall chirality. We hope to report on some of the above issues in the near future.\\

{\it The present investigation is theoretical and the manuscript has no data to be associated with.}\\

%%%%%%%%%%%%%%%%%
\acknowledgements{This work is supported by the Brazilian agencies Coordena\c{c}\~ao de Aperfei\c{c}oamento de Pessoal de N\'ivel Superior (CAPES), grant No.~88882.440276/2019-01 (MP), by Conselho Nacional de Desenvolvimento Cient\'ifico e Tecnol\'ogico (CNPq), grant No. 303469/2019-6 (DB), by Paraiba State Research Foundation (FAPESQ-PB) grant No. 0015/2019, and by Federal University of Para\'\i ba (UFPB/PROPESQ/PRPG) project code PII13363-2020.}

\end{document}